\documentclass[aps,prd,twocolumn,floats,nofootinbib,longbibliography]{revtex4-1}
\usepackage[dvips]{graphicx}

\usepackage{amsmath}
\usepackage{braket}
\usepackage{amsfonts}
\usepackage[colorlinks=true]{hyperref}
\usepackage{hyperref}
\usepackage{xcolor}

\usepackage{booktabs}

\usepackage{caption}
\usepackage{graphicx}
\usepackage{subcaption}

\usepackage{bm}

\def\aap{\textit{Astronomy and Astrophysics}}

\def\be{\begin{equation}}
\def\ee{\end{equation}}
\def\bea{\begin{eqnarray}}
\def\eea{\end{eqnarray}}

\begin{document}
\title{Strong lensing in multimessenger astronomy as a test of the equivalence principle}

\author{Olivier Minazzoli}

\affiliation{Centre Scientifique de Monaco, 8 Quai Antoine 1er, 98000, Monaco}
\affiliation{Artemis, Universit\'e C\^ote d'Azur, CNRS, Observatoire C\^ote d'Azur, BP4229, 06304, Nice Cedex 4, France}
\begin{abstract}
Standard Shapiro delay-based test of the equivalence principle, which are grounded on the measurement of two arrival times from a unique source but from messengers with different properties, cannot produce a reliable quantitative test of the Einstein equivalence principle. Essentially because they are based on the estimation for different messengers of the one-way propagation time between the emission and the observation that is not an \textit{observable} per se. 
As a consequence, such tests are extremely model dependent, at best. In what follows, I argue that the differential arrival times for strongly lensed multimessengers can be used to define a new test of the Einstein equivalence principle that is both well-defined from a relativistic point of view and model independent---because it is entirely based on actual \textit{observables}.

\end{abstract}
\maketitle

\section{Introduction}
The standard Shapiro delay-based test of the equivalence principle relies on the measurement of two arrival times from a unique source but from messengers with different properties\footnote{Which are nevertheless expected to propagate at (or nearly) the speed of light \cite{krauss:1988pl,longo:1988pl,pakvasa:1989pr}---see \cite{minazzoli:2019pr} for a list of studies based on this approach.}, and from the assumption that one can estimate the time of flight of the messengers from indirect observations and a well-suited space-time model. It is meant to test whether or not various messengers propagate along the same trajectories, as it is expected from the Einstein equivalence principle---which encompasses notably both the weak equivalence principle and the Lorentz invariance \cite{will:1993bk}. In its most common form, it is based on the assumption that the universe is flat at the relevant scale, and that the Shapiro delay $\delta T $ with respect to the propagation time that would have happened in a Minkowski spacetime is given at leading order by
\be
\label{eq:gen_usual_shapiro}
\delta T = - \frac{1+\gamma}{c^3} \int_{\bm r_E}^{\bm r_O} U(\bm{r}(l)) dl ,
\ee
where $\bm r_E$ and $\bm r_O$ denote emission and observation positions, respectively, $U(\bm{r})$ is the Newtonian gravitational potential, and the integral is computed along the trajectory. $\gamma$ is the parameter appearing in the space-space component of the $c^{-2}$ parameterized post-Newtonian metric \cite{will:1993bk}. In the parameterized post-Newtonian framework, $\gamma$ will be the same for different messengers, since it is a property of the metric. In the standard test of the equivalence principle, the parameter $\gamma$ is used as a proxy to quantitatively infer at what level the Shapiro delay is indeed the same for different types of messengers, but is not derived from a fundamental theory---unlike the usual parameterized post-Newtonian parameters, wich always have a one-to-one correspondence with parameters or functions of a given theory \cite{will:1993bk}.

In practice, from the measured delay between the arrival times of the two messengers ($X$ and $Y$) $\Delta \tau_a := \tau_a^X - \tau_a^Y$ and the assumed delay between their emission times $\Delta \tau_e := \tau_e^X - \tau_e^Y$, the difference between the parameters $\gamma$ for different messengers would be infered with the equation that follows
\be
\label{eq:gammaOld}
\gamma_X-\gamma_Y = c^3\frac{\Delta \tau_e-\Delta \tau_a}{ \int_{\bm r_E}^{\bm r_O} U(\bm{r}(l)) dl}.
\ee

However, as explained in \cite{gao:2000cq}, Eq. (\ref{eq:gen_usual_shapiro}) is gauge (or background) dependent, in the sense that a coordinate change that preserves the Minkowski background can modify even the sign of the numerical value of the effect. This is due to the fact that there is no meaningful way to compare the propagation times that would arise in a flat and a curved spacetime in general \cite{gao:2000cq}. Hence, there is no meaningful definition of what a delay with respect to an hypothetical propagation time in a Minkowski spacetime is; while this is precisely what Eq. (\ref{eq:gen_usual_shapiro}) is about. 

Nevertheless, it has been shown in \cite{minazzoli:2019pr} that something that effectively reduces to the usual formulation of the Shapiro delay can be defined in a gauge independent way---albeit an observer dependent one---as the difference between the total propagation time $T$ (which is still not an observable in general) and the luminosity distance $d_\text{lum}$ over the speed of light ($\delta T = T - d_\text{lum}/c$).
It recovers the standard Shapiro delay equation (\ref{eq:gen_usual_shapiro}), when one assumes a post-Newtonian metric, for observers at rest with respect to the background, up to higher-order effects, and provided that the magnification from gravitational lensing is negligible.

But the standard use of Eq. (\ref{eq:gen_usual_shapiro}) also suffers from an implicit assumption that there exists a coordinate time such that the potential and its derivative vanishes at infinity. In \cite{minazzoli:2019pr}, it was  shown that such an assumption leads to a spurious divergence of the Shapiro delay with the number of sources, which however gets renormalized by simply tying the coordinate time to an observer's proper time instead---such that the potential and its derivative vanish at the location of the observer; and not at infinity. The fix has the additional advantage that it depicts the effect directly in terms of an observer's proper time. However, with the use of an adequate coordinate time, it was shown that the Shapiro delay is no longer monotonic with respect to the increase of sources, because the sign of the effect becomes dependent on the geometrical configuration of the propagation with respect to a source. Hence, it was shown in \cite{minazzoli:2019pr} that any estimate of the Shapiro delay is extremely sensitive to the incompletness on the knowledge of gravitational sources, whereas all catalogs are incomplete by construction. The sign of the effect for the transient GW170817 has even been shown to depend on the catalog of galaxy clusters being used for the estimation.\footnote{Note also that, in some situations, the Shapiro delay can be arbitrarily close to $0$ depending on the geometrical configuration of the propagation with respect to the gravitational sources---see Fig. 3 in \cite{minazzoli:2019pr}.} 
As a consequence, all analyses based on Eq. (\ref{eq:gen_usual_shapiro}) that aimed to test the equivalence principle on cosmological scales got their inference of the amplitude of the Shapiro effect wrong, and therefore their inference of the quantitative constraints on the equivalence principle wrong as well. Not to mention that Eq. (\ref{eq:gen_usual_shapiro}) entirely neglects cosmology \cite{minazzoli:2019pr}.

Another test has recently been suggested, based on the idea that there should be an additional delay induced by the bending of light \cite{rubin:2020ar}, which ougth to be the same for different messengers if the Einstein equivalence principle is correct. However, likewise, given that this test is based on the non-observable one-way propagation time, the computation of the effect reduces, in essence, to an hypothetical comparison between the propagation of light in a flat and a curved space-time. But, any such comparison is coordinate dependent, as follows from \cite{gao:2000cq}. 

The fact that messengers with different properties are witnessed to arrive at the same time is an interesting evidence of the validity of the Einstein equivalence principle, but it can only be a qualitative evidence as one cannot derive a reliable quantitative test from a relativistic effect that is not an observable per se---here, the time delay(s), or advance(s), of the one-way time of flight with respect to what would have happened in an hypothetical Minkowksi spacetime. What can make sense from the relativistic point of view, on the other hand, is the difference between the propagation times from different trajectories in the (unique) spacetime that depicts our universe. This is notably what Shapiro-delay-based tests in the Solar System rely on, e.g. \cite{bertotti:2003na}: a differential measurement of the propagation time; and not a comparison of a propagation time with respect to a fiducial propagation in an hypothetical flat spacetime.

\section{A novel and rigorous test}

Strong lensing with multiple messengers from a same source allows one to define a rigorous test of the Einstein equivalence principle, as one can construct a test that uniquely relies on observables. Indeed, irrespective to any theoretical model, a quantitative measure of how close trajectories for different messengers are 
 is simply given by
\be
\label{eq:newtest}
\xi_{XY} =  \frac{\Delta t_{i j}^X - \Delta t_{i j}^Y}{\braket{\Delta  t_{i j}}},
\ee
where $\Delta t_{i j}^Z$ is the measured arrival delay between the two lensed images ($i$ and $j$) of the same source for a messenger of type $Z$, and $\braket{\Delta  t_{i j}}$ is the mean of the two delays involved. $\xi_{XY} \neq 0$ would signal a violation of the Einstein equivalence principle if the messengers were expected to follow the same trajectories due to this principle. Less generally, one still could use the usual post-Newtonian-inspired parametrization of Shapiro delay-based tests, but now applied to the whole time delay $\Delta t_{i j}$ between two lensed images ($i$ and $j$) of the same object---i.e. $\Delta t_{i j} \rightarrow (1+\gamma) \Delta t_{i j} /2$, see for instance Eq. (84) in \cite{linet:2016pr}. Indeed, for models that would predict different $\gamma$ for different messengers, the conversion in terms of $\gamma$ simply reads $\gamma_X-\gamma_Y = 2~\xi_{XY} $, as long as the violation of the equivalence principle is small (i.e. $|1-\gamma_Z| \ll 1 ~\forall Z$). 

Let us now stress that inferring $\xi_{XY} $ only relies on the measurements of the two pairs of arrival times. In particular, one does not need to know anything on physics behind the emissions of the different messengers, nor to model the lens, nor the background metric, nor even the fundamental way through which the propagation is affected by the background metric. Unlike Eq. (\ref{eq:gammaOld}) that relies not only on an assumed emission time delay (which depends on the uncertain physics of the emissions), but also on a specific space-time model, its coupling with matter fields, and an indirect evaluation of the potential along the propagation of the messengers; Eq. (\ref{eq:newtest}) is auto-sufficient and is obviously a quantitative measure of how similar would be the trajectories of different messengers from the same source. As a consequence, Eq. (\ref{eq:newtest}) is undoubtedly well suited to test the Einstein equivalence principle for any astrophysical or cosmological model, for any source, and for any theory. One just need messengers from the same source that are expected to travel along the same trajectories following the Einstein equivalence principle.\footnote{For instance, for a comparison involving gravitational and electromagnetic waves, one has to make sure that the propagation of the former follows the laws of geometric optics at the relevant frequencies, which is not always the case \cite{takahashi:2017aj}.}

Now note that an equation similar to Eq. (\ref{eq:newtest}) has been suggested in the literature already in \cite{yu:2018:ec}. However, this other equation remains model dependent, while it really does not need to be. Indeed, $\xi_{XY}$ is a measure of potential deviations from the equivalence of the propagation of different messengers. If the messengers were expected to follow the same trajectories, evidence of a non-null $\xi_{XY}$ indicates a violation of the Einstein equivalence principle. The introduction of a model dependency is facultative. Nevertheless, any measurement of $\xi_{XY}$ can be associated to the location of the lens on the celestial sphere, in order to encompass potential directional-dependent violations, as one could expect from theories that break the Lorentz invariance for instance.

\section{Conclusion}

To conclude, now that we fully entered the era of multimessenger astronomy, it may not last too long before we have the opportunity to make the observations for two different messengers of the arrival delay between two lensed images of the same source---like, e.g., following the merger of a binary neutron star \cite{burns:2019ar}, with the conjoint observation of strongly lensed gravitational waves and gamma rays. As argued in this manuscript, it will provide a novel and rigorous test of the Einstein equivalence principle.\\

\begin{acknowledgments}
The author thanks Eric Burns for some of his suggestions on the manuscript, and also for finding out the reference \cite{yu:2018:ec}, as well as Nelson Christensen for carefully reading the original version of the manuscript. This proposal follows from many discussions, in particular with Nathan Johnson-McDaniel, Mairi Sakellariadou, CS Unnikrishnan, Daniel Holz, Robert Wald and many more, notably within the LIGO-Virgo collaboration.
\end{acknowledgments}

\bibliography{Slensing}

\end{document}